\documentclass[epj]{svjour}
\usepackage{latexsym}
\usepackage{graphics}
\begin{document}

\title{
Bloggers Behavior and Emergent Communities in Blog Space 
} 
\author{Marija Mitrovi\'c \and Bosiljka Tadi\'c}%
%
\institute{Department  for Theoretical Physics, Jo\v{z}ef Stefan Institute, 
P.O. Box 3000, SI-1001 Ljubljana, Slovenia
}
\date{Received: date / Revised version: date}
\abstract{
Interactions between users in cyberspace may lead to phenomena different from those observed in common social networks. Here we analyse large data sets about users and Blogs which they write and comment, mapped onto a bipartite graph. In such enlarged Blog space we trace user activity over time, which results in robust temporal patterns of user--Blog behavior and the emergence of communities. With the spectral methods applied to the projection on weighted user network  we detect clusters of users related to their common interests and habits.  Our results  suggest that different mechanisms may play the role in the case of very popular Blogs. Our analysis  makes a suitable  basis for theoretical modeling of the evolution of cyber communities and for practical  study of the data, in particular for an efficient search of interesting Blog clusters and further retrieval of their contents by text analysis.
\PACS{
      {89.75.Hc}{Networks and genealogical trees}   \and
      {89.75.Fb}{Structures and organizations in complex systems}\and
      {02.70.-c}{Computational techniques; simulations}
     } 
}
\titlerunning{Bloggers and Blogs Communities}
\maketitle
\section{Introduction\label{sec-intro}}
In the information society, getting information from immediate users becomes an important way in opinion making, marketing products, and in everyday life issues as health, hazards, traffic problems, etc.  In this respect the contemporary  communication networks and social media offer  on-line interactions, which facilitate communications, but on the other hand, they introduce a new kind of technology-mediated social clustering not known in the history. Behavior of users in the cyber space may be altered in comparison with common social interactions, which may induce new phenomena in the fast developing technology-based society. 
Obvious  ``absence of person'' (the communication is indirect, mediated by text or video material posted on the web portals), accessibility, massive data available, and fast communications are some of the reasons which affect blogging at microscopic scale. The collective behaviors emerging in these interactions have not been yet understood.   Therefore, scientific data analysis and modeling from the point of view of the complex evolving systems appears as a necessary step towards better understanding of the social phenomena in cyberspace.

Among variety of the currently available techno-social organizations, Blogs are somewhat specific,  between the ordinary Web pages, where the owner writes its contents, and much faster information exchange between friends in  FriendFeed, Facebook or MySpace networks. On the other hand, on Blogs the posts are  short written forms, usually dedicated to a given subject and written by known author (or a group of authors) of the Blog. Other registered users can read and leave comments on the posts. Registration of users is required at many Blog sites and the action of each user is traced in time. The action of users is delayed compared with posting time. Apart form the subject, the act of writing the post and/or comment often requires certain dedication and writing skills, in which author's personal profile and preferences get involved. In this way, posts may comprise an aesthetic as well as emotional and moral contents, similar to books, movies or music items accessible on the Web. Unlike friends networks, in-advance relationship between authors of Blogs and other users is not an underlying  context on Blogs.  This all might influence the evolution of Blogs, however, no precise measure of the importance of these and other factors exists so far. The mechanisms that drive user activity in the Blog space have not been fully understood \cite{nature09}. 

Analysis of data in different social media  revealed  correlated behavior manifested in power-laws in the structure of networks and communities related to movie \cite{grujic2009,lorenz2008} and music  genres \cite{lambiotte2005,lambiotte2006}, and cascades of events in the Internet forums \cite{kujawski2007} and through different  Blogs via  hyperlink connections between them \cite{bachnik2005,fu2006,leskovec2007,liu2007}.    Further regularities were found in the anlysis of the Blog entries \cite{sano2009,mike2007}. The approach with a targeted analysis of the contents of posted text on Blogs and in other forms \cite{thelwall2009b,fortuna2007}, brings a new dimension in the study of the techno-social communications \cite{science2009}. 

In this work we consider the data about actual events occurring between users-and-posts and map them to a {\it bipartite network}, in which users, as one partition, do not have a direct connection but interact through the posts, which makes the other partition. Links between users and posts are also directed, depending on the action (reading or writing the post/comment). In this way, our network makes an enlarged Blog space, with users and posts (and comments to the posts) treated at the equal level. It is also related to a given Blog site (we consider two Blog sites with different organizational characteristics). In this enlarged Blog space we are able to study the post-mediated interactions between users and how the behavior of users affects the structure of the Blogs. 
Note that in  several previous studies of Blogs \cite{bachnik2005,leskovec2007,liu2007} different networks have been considered: posts as the network nodes were connected with the hyperlinks pointing from one post to another, as the network edges.

Organization of the paper is as follows: In Sec.\ \ref{sec-tempo} we analyse the temporal features of both users and posts and determine statistical measures of users activity relevant to the Blogs. In sec.\ \ref{sec-binet} we define the bipartite networks as emanating from the data and study their topological properties. In Sec.\ \ref{sec-communities} we determine the community structure of users and fine clustering of  posts within user groups. The difference between very popular and other posts is also pointed at the level of such communities. A short Summary of the results and conclusions are given in Sec.\ \ref{sec-Summary}.

\section{Temporal features of blogging\label{sec-tempo}}

\subsection{Data Structure}
We consider large Blog data, together about half of a million entries, collected  from two different Blog sites,  which have entirely different internal organization and history: BBC Blogs and  Belgrade radio B92  Blogs. The case of B92 Blogs is interesting for the analysis for several reasons. First, we have collected all data from the beginning 27. May 2007 till 1. March 2009.
Furthermore,  on this Blog site users are registered not just to read and comment other posts, but to write their own post, and more importantly, no predefined categories of post subjects are imposed. Thus, the internal structure of posts emerges in a self-organized manner through user interactions on posts and comment-on-comment actions. The availability of posts is time limited to seven days 
(this rule was imposed after first few months of the functioning). Some of the users are upgraded to so called VIP authors, whose number fluctuates in time, and their recent posts are highlighted. Here we analyzed the posts written by all VIP users in the above mentioned period and consider {\it all users related to these posts},  which comprises of $N_U=4598$ users, and $N_B=4784$ posts  and $N_c=406527$ comments to these posts. 
On the other side, the BBC Blogs exists for much longer time. For better comparison, we have collected the data form the same period as above, which gives $N_B=3792$ posts, and $N_c=80873$ comments written by $N_U=21462$ registered users. In contrast to B92 Blogs, at the BBC Blogs both authors of posts  and category of posts are predefined and fixed. The users are registered and allowed only to read and comment the posts.  Accessibility of posts is not limited in time. 
In contrast to B92 Blogs,  information about ID of a comment which is commented is not available in BBC Blogs, all comments are attributed to the original post.

\subsection{User behavior in Blog Space}
\begin{figure}
\centering
\begin{tabular}{cc}
\resizebox{22pc}{!}{\includegraphics{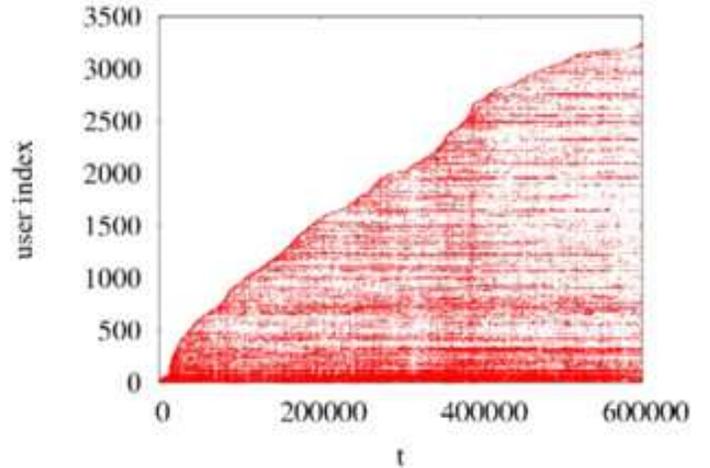}}
\end{tabular}
\caption{Temporal linking pattern of users (ordered by time of first appearance) within roughly first year since the opening of the B92 Blogs. }
\label{fig-userpattern}
\end{figure}

\begin{figure}
\centering
\begin{tabular}{cc}
\resizebox{20pc}{!}{\includegraphics{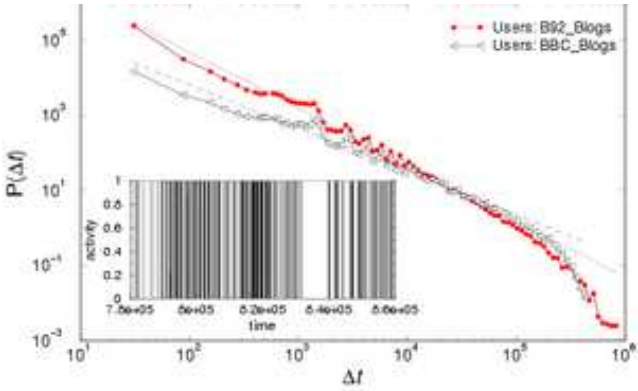}}\\
\end{tabular}
\caption{Distributions of time intervals in users activity on B92 Blogs  and on BBC Blogs, averaged over all users of given Blog site. Inset: Example of 'bar-code' of a very active user on B92 Blogs.}
\label{fig-pdt}
\end{figure}

Data about users, kept under their registered IDs, contain information about users' activity over time, appearance of their posts, and their comments to other posts. In Fig. \ \ref{fig-userpattern} we show temporal patterns of the activity of more than 3000 users, ordered by the time of their registration on B92 Blogs within first year. As the figure shows, users appear in some waves, probably related to the external events or Blog site management, and newly registered users are active within some time intervals, whereas their activity is reduced in later times. Some users persist over long time period, while many other users either reduce frequency or stop writing on Blogs altogether. The heterogeneity in the users' activity is further quantified by  the analysis of time intervals $\Delta t$ between two successive user's activities and  by the number of events  (written posts and/or comments) $n^U_{com}(i,t)$  within a specified time window $T_{WIN}$, for instance within one day.  The distribution of time intervals between user's successive activities is given in Fig.\ \ref{fig-pdt} for both BBC and B92 Blog users. The power-law dependences over several decades (time is measured in minutes) suggest robust non-random patterns of users behavior, which is not much dependent on how the Blog site is organized. The slopes of the curves are different (1.5 and 1.15 at B92 and BBC Blogs, respectively), indicating slightly larger probability of large inactivity times at BBC Blogs.


\begin{figure}[h]
\centering
\begin{tabular}{c}
\resizebox{18pc}{!}{\includegraphics{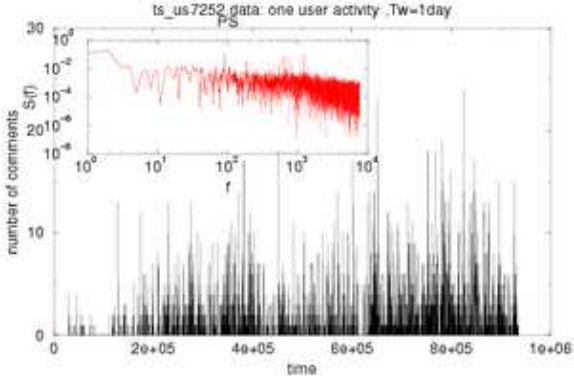}}
\end{tabular}
\caption{ Time series of an active user  on B92 Blogs and (Inset:) its power spectrum.}
\label{fig-userts}
\end{figure}

Further quantitative analysis of the time series of  activity, $n^U_{com}(i,t)$,  a given user $i$ in time window $t$,  reveals additional information about user behavior. An example of such time series of a very active user from B92 is shown in Fig.\ \ref{fig-userts}, with $T_{WIN}$ equals one day. The power-spectrum of the time series shows long-range correlations at large frequency region, which suggests that the user activity is correlated over small time intervals.

\begin{figure}[h]
\centering
\begin{tabular}{c}
\resizebox{18pc}{!}{\includegraphics{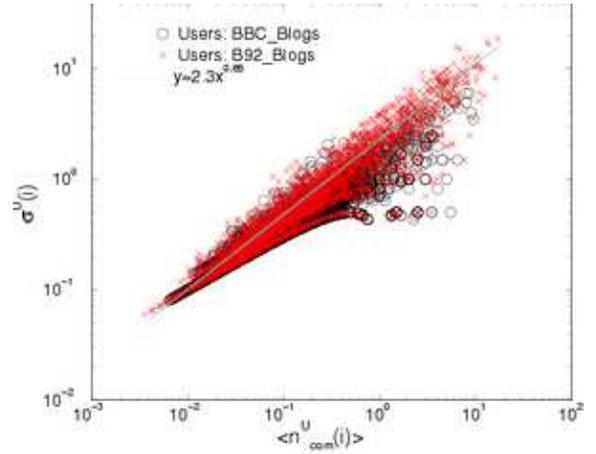}}\\
\end{tabular}
\caption{Scaling of the dispersion  $\sigma ^U(i)$ of users activity  (time bin one day) time series   plotted against its average value $\langle n^U_{com}(i)\rangle$  for all considered users  on B92 and on BBC Blogs. }
\label{fig-sigmahusers}
\end{figure}
We also analysed such time series for all users. In Fig.\ \ref{fig-sigmahusers} the dispersion of these time series $\sigma ^U(j)$ of a given user $j$ is plotted against its  average over all time windows $<n^U_{com}(j)>$. Thus in this plot each point represents one user from our list. The plot obeys the scaling relation
\begin{equation}
\sigma ^X(i) = c <n^X_{com}(i)>^\mu \ ,
\label{eq-sigmah}
\end{equation}
which is found in many complex dynamical systems from social to biological \cite{eisler2008,zivkovic2006}. It is interesting that all users in the considered Blogs on both BBC and B92 Blogs follow the same scale invariance, with the exponent $\mu \sim 0.88$. Note that the largest exponent $\mu =1$ is generally expected in strongly driven dynamical system, whereas the lower limit $\mu =1/2$ is found in random (uncorrelated) events \cite{eisler2008}. The observed scale invariance of the user time series in Fig.\ \ref{fig-sigmahusers} strongly indicates non-randomness in the user activity on Blogs.

\subsection{Temporal Patterns in Blogs}

The fractal   temporal pattern of users activity is manifested on events recorded at Blogs, in particular, each user action results in either a new post or a new comment to some of the existing posts. From the data on B92 Blogs we can trace where each particular user action was directed to, by analysis of the temporal patterns of the related posts. 

The question of 'How long it takes to people to react to new posted material?' 
was studied in other communication systems, e.g., e-mails, Youtube videos, donation to tsunami victims etc, see   \cite{crane2009} and references therein. The appearance of the power-law distribution of the response times $P(t-t_i)$ to an event $i$ posted at time $t_i$ was associated with human nature of acting with priority queues \cite{grinstein2008,crane2009}. Assuming a single-server-queue limit and random arrival of  events to the queue, in ref.\ \cite{grinstein2008} a universal power-law distribution of the waiting (or response) times was derived with the exponent 3/2 corresponding to the situation when the average arrival rate exceeds the average execution rate. An exponent larger than $2$ is expected in the opposite situation, as discussed in Ref.\  \cite{crane2009}, where attention rather than priority was stressed as a key mechanism. In both cases, a theory of independent queues were considered. Note that in the network environment, where the queues are mutually interacting, for instance in the packet queuing processes \cite{tadic2004} with LIFO queue, the waiting time distribution exhibits power-law with an exponents depending of the traffic density, dropping below 2 when the jamming on the network occurs. 

\begin{figure}
\centering
\begin{tabular}{c}
\resizebox{18pc}{!}{\includegraphics{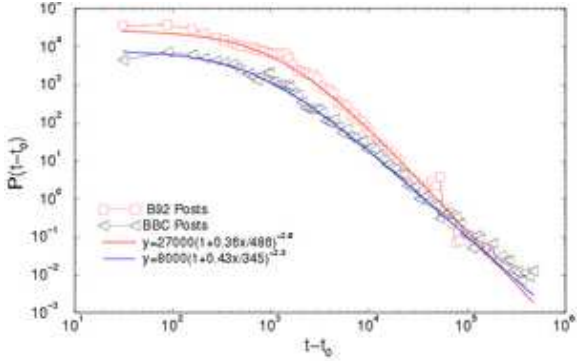}}\\
\end{tabular}
\caption{Distribution of the response time to posts on B92 and BBC Blogs. Fit curves by the $q$-exponential form Eq.\ (\ref{eq-qexp}).
}
\label{fig-Blogsresptime}
\end{figure}

In the Blog data we construct the distributions $P(t-t_i)$ having the posting time $t_i$ of all posts and  the action time $t$, i.e., posting a comment to the post $i$. The distributions averaged over all posts are shown in Fig.\ \ref{fig-Blogsresptime} for both B92 and BBC Blogs, with slopes 2.8 and 2.3 respectively. Again, it's remarkable that the power-law tail in both cases exists, which can be fitted with the $q$-exponential expression
\begin{equation}
P(t-t_i) = C\left(1-(1-q)\frac{t-t_i}{t^*}\right)^{\frac{1}{1-q}} \ .
\label{eq-qexp}
\end{equation}

The differences in the exponents might be attributed to closing the posts in B92 after a preset expire date, which is not the case with the posts on BBC Blogs.
In Ref.\ \cite{crane2009} an exponent larger than two is expected when the subject is repeatedly brought to people's attention, which also might be the case with the highlighting recent (or very active) posts on the Blogs.

\begin{figure}
\centering
\begin{tabular}{cc}
\resizebox{20pc}{!}{\includegraphics{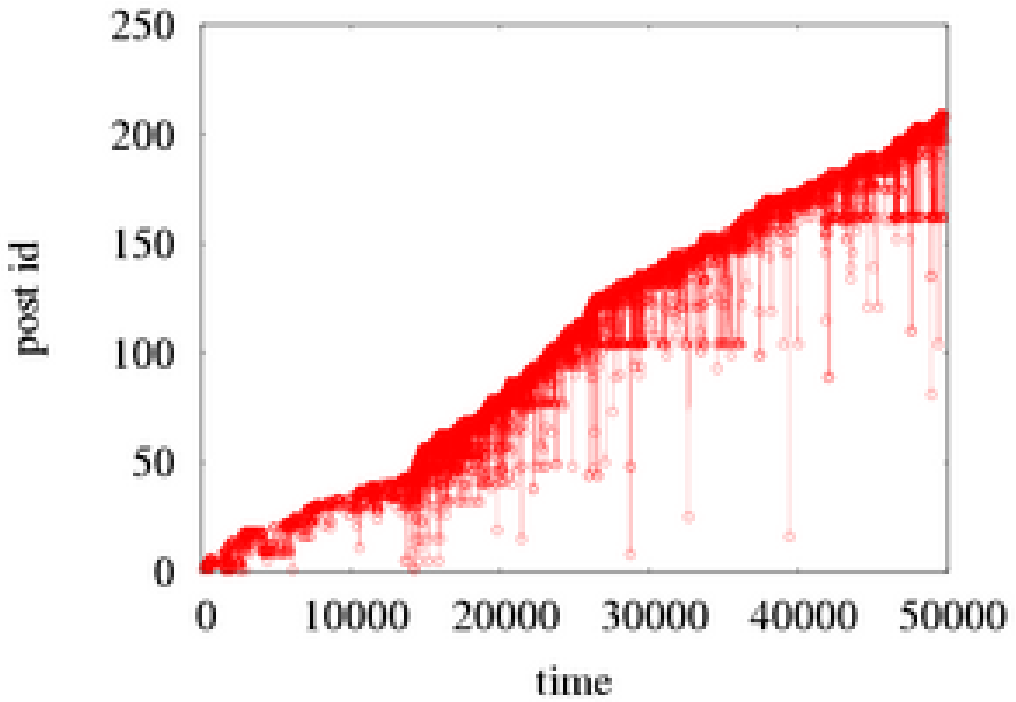}}\\
\resizebox{20pc}{!}{\includegraphics{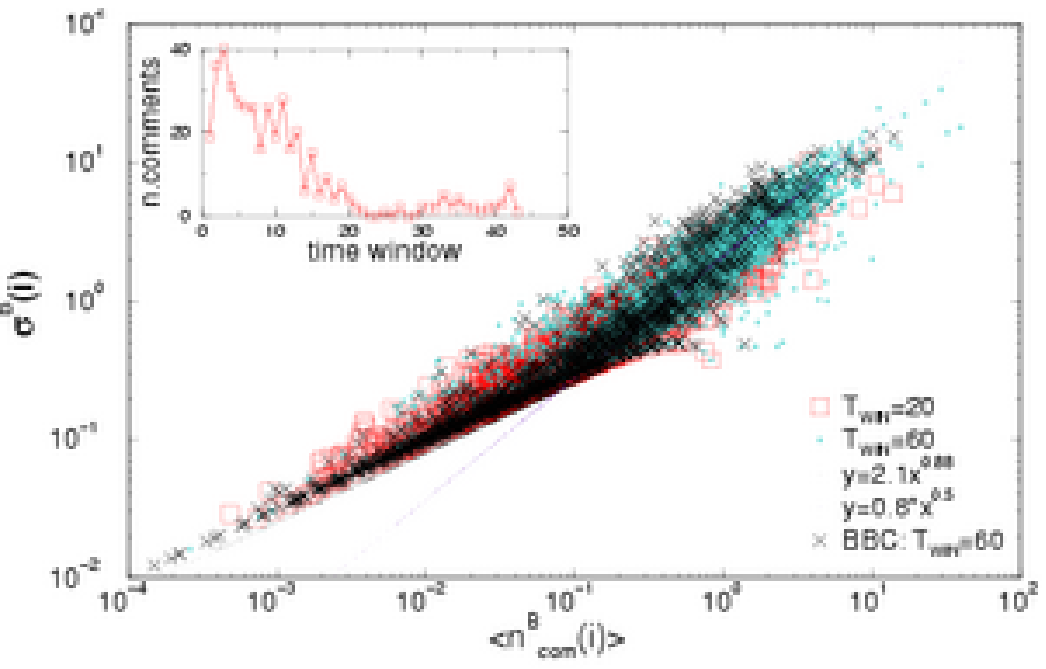}}\\
\end{tabular}
\caption{ (top) Linking pattern over times on B92  posts.  (bottom) Dispersion vs average for posts with time bin 20 min and 1 hour for B92 and for  BBC Posts with 1 hour bin.}
\label{fig-Blogsigmah}
\end{figure}

The temporal pattern of linking to a particular post  is shown in Fig.\ \ref{fig-Blogsigmah}(top) for first 250 posts on B92 Blogs (no expire dates of posts was imposed at that time). It shows that posts becomes less interesting after certain period of time. Similar pattern was observed in the case of BBC Blogs. Accordingly, the time series consisting of the number of comments (within a time bin) on a  typical post  shows an increase shortly after the post appearance, then it drops to a steady value and eventually stops, an example with time bin of 1 hour is shown in the inset to Fig.\ \ref{fig-Blogsigmah}(bottom). 
The scaling plot of all posts time series, similar to the one of users in Fig.\ \ref{fig-sigmahusers}, is shown in Fig.\ \ref{fig-Blogsigmah}(bottom). In contrast to users, here one can see that posts belong to two distinct categories according to the popularity and, consequently, their dispersion of the time series are different. The active posts show large average number of comments per time bin and at the same time large dispersion of the time series according to Eq.\ (\ref{eq-sigmah}) with the exponent $\mu =0.68$. Whereas a large group of posts both in BBC and B92 Blogs exhibits a random variation in the time series, these posts are represented by the lower-left part of the plot, where the exponent is $\mu=0.5$.
The behavior is statistically similar on both Blog sites, with the exception of few very active posts in the B92 Blogs (appearing close to the tip of the plot).

\section{Blog data mapped onto bipartite network\label{sec-binet}}

\begin{figure}
\centering
\resizebox{18pc}{!}{\includegraphics{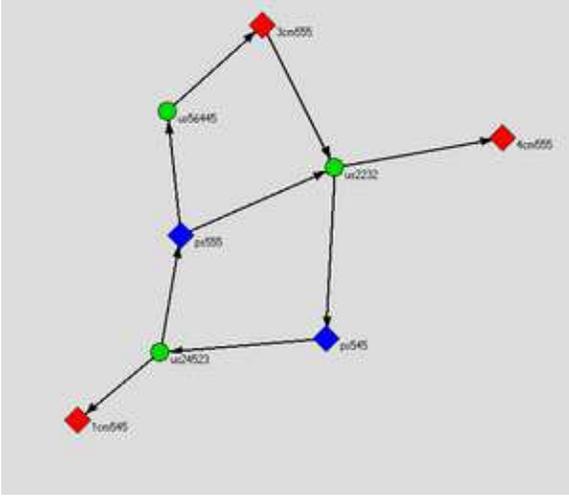}}
\caption{ Example of bipartite network of blogs and users. Users, shown as green circles, are connected with posts, shown as  blue diamonds, and to comments on posts, red diamonds, but not among themselves. }
\label{bipartite_net}
\end{figure}

As mentioned in the Introduction, in our approach the data about users and Blogs are mapped onto a bipartite network, with {\it users as one partition} and {\it posts and comments, as the other partition}. Here we explain how the mapping is exactly done and what kind of networks emerge. One should stress that, by definition, in the bipartite networks no direct link occurs between the nodes within the same partition. Thus two users in our network interact with each other only through the posts that they write and/or chose to comment.
A directed bipartite network is most suitable to represent the events contained in the data. In the data we have $i_U=1,2, \cdots N_U$ users and $j_B=1,2, \cdots N_B$ posts and comments, which make the bipartite network nodes.  The links between users and posts (or comments) are inserted  as follows:  

\begin{itemize}
 \item a directed link from the user $i_{U}$ points to the post $j_{B}$ which that user posted; Similarly a link  $i_U \to mcmj_B$ from user $i_U$ to $m$th comment of the post $j_B$ is drown when the data indicate that the user $i_U$ wrote that comment; 
 \item a link from the post $\ell _B$ to the user $k_U$ is drown when the data indicate that the user red that post and/or one of its comments. In addition, a link points from $r$th comment of the post $\ell _B$ towards that user, $rcm\ell_B \to k_U$, if the user red that comment. Note that the indication that a user red a  post/comment is contained in that user comment's data. 
\end{itemize}
The mapping is illustrated in Fig. \ref{bipartite_net}.
Note that in this mapping multiple links are possible, e.g., if the user replied to comments on his post or comments. 

\subsection{Statistical properties of bipartite graphs}
One of the immediate consequences of the above explained linking pattern over time is the inhomogeneous network of users and Blogs. Here we  characterized it by two measures: the degree distribution of each partition and by the number of commons: the common posts per pair of user and vice versa, which are relevant for the community structure further analysed in this paper.

\begin{figure}
\centering
\begin{tabular}{cc}\resizebox{22pc}{!}{\includegraphics{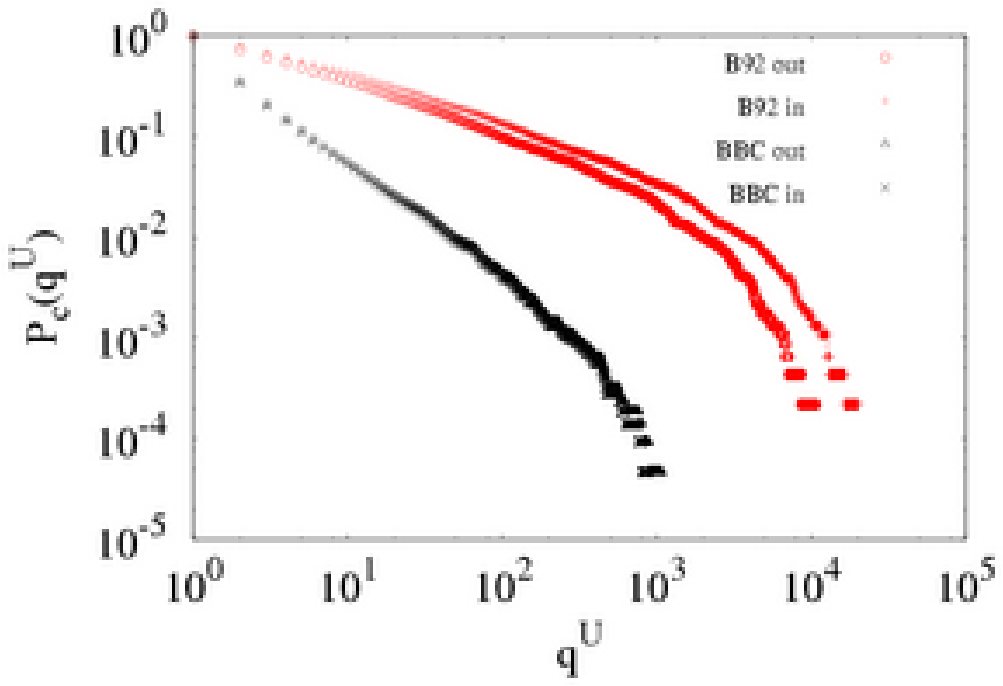}}\\
\resizebox{22pc}{!}{\includegraphics{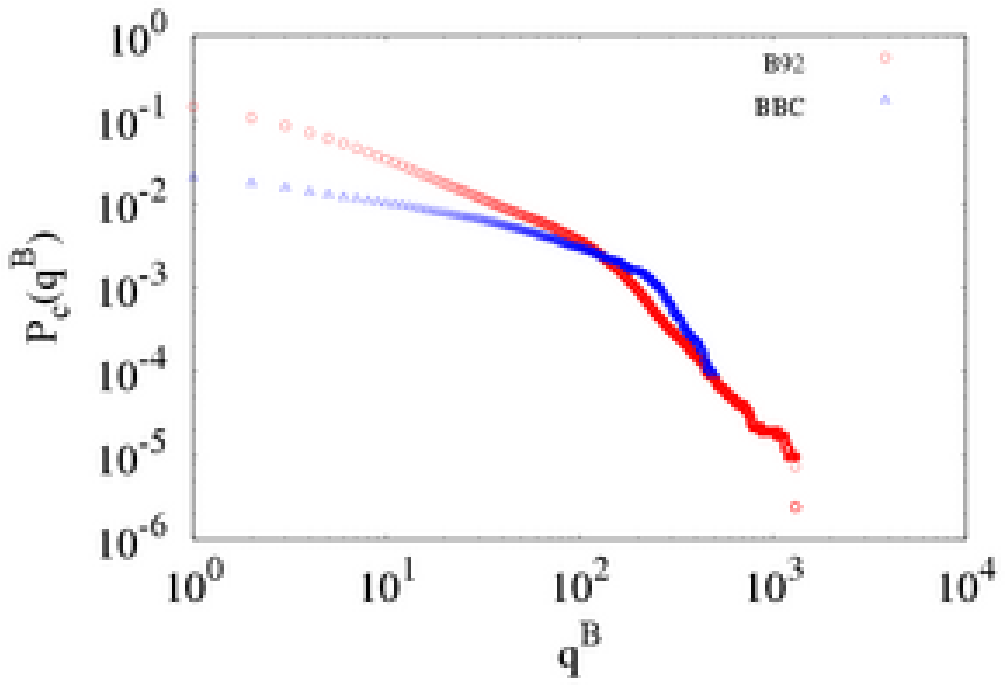}}\\
\end{tabular}
\caption{ Bipartite network analysis:  Cumulative distributions for in- and out-degree of users (top) and out-degree of posts (bottom) for data from B92 and BBC Blogs.
}
\label{fig-degree}
\end{figure}

\textbf{Degree distribution} in the case of our directed bipartite network can be determined for in- and out-degree and in both partitions. For the user partition, the out-degree is defined as the number of posts and comments written by the user, while in-degree is the number of posts and comment red by that user. Note that in the structure of B92 Blogs, where the users are allowed to write new posts and where also comment-on-comment is allowed and notified in the data, in- and out-degree distributions are different, as shown in Fig.\ \ref{fig-degree} (top panel).  In the BBC Blogs, however, the users are only allowed to write comments on the original posts, the difference between in- and out-degree distribution of users entirely disappears, see Fig.\ \ref{fig-degree}. In te partition of of posts and comments, also shown in Fig.\ \ref{fig-degree} (bottom panel), in-degree of posts is  equal to one (each post or comment have one author), while  the out-degree of posts has a nontrivial distribution. Apart from the multiple linking mentioned above, the out-degree of a post is roughly equal to the number of users who wrote a comment (including comments on comments) of that post. Again, there is difference in the slope of the decay of this distribution for the BBC Blogs data and B92 Blogs data. However, the qualitative features are similar: a power-law decay for less popular posts (with number of comments smaller than a characteristic value $n_{com}^*$ (approximately 100-200). In both cases a sharp bending of the distribution occurs for the popular posts, on which the number of comments exceeds  $n_{com}^*$.  As mentioned above, the slopes of the power-law distributions are not universal, here we mention them for completeness. The cumulative distributions are shown in Fig.\ \ref{fig-degree}, the exponents are compatible with the differential degree distributions as follows: 1.57 and 2. for  the user out-degree, and  1.98 and 1.5 for the posts out-degree in B92 and BBC Blogs, respectively. 

\begin{figure}
\centering
\begin{tabular}{cc}
\resizebox{22pc}{!}{\includegraphics{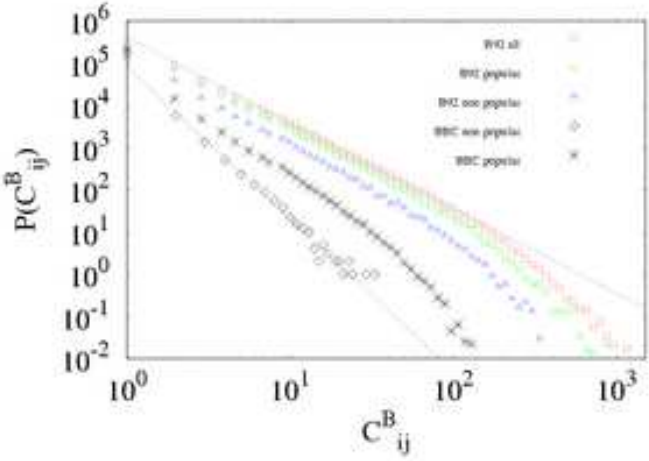}}\\
\end{tabular}
\caption{Distribution of commons: Number of common posts  $C_{ij}^B$  per pair of users for B92 and BBC Blogs, as indicated. }
\label{fig-commons}
\end{figure}
\textbf{Distribution of Commons} is another measure emanating from the bipartite representation of the network, and determines weight of links in a suitable mono-partite projection. One can consider projections in both partitions \cite{grujic2008}. Here we are interested in the behavior of users in the enlarged Blog space. Therefore we will consider the projection on the user partition, where the commons, $C_{ij}^B$ are defined as {\it the number of common posts and comments per pair of users}.  
Distribution shown in Fig.\ \ref{fig-commons} of commons $C_{ij}^B$ per pair of users obtained from the whole dataset, non-popular, and popular posts, as defined above, for B92 Blogs.  
The distribution has characteristic power law decay for all sets of data. Similar features are found in the BBC Blogs data, but with an order of magnitude  smaller cut-off. This also indicates that considerably larger  overlap between users occurs on B92 Blogs compared to BBC Blogs, which can be attributed to much larger freedom in the authorship and the subject categories on B92 Blogs.

\section{Community Structure in Blogs\label{sec-communities}}
The analysis of commons discussed above helps us to trace the (mediated) interactions between users in the Blog space, which is of our primary interest in this paper. In particular, by projecting the bipartite graph onto monopartite user space, the common number of posts  $C_{ij}^B$ for the pair $\{ij\}$ of users appears as a weight on the links in the user subgraph. In weighted networks communities can be detected both according to the topology, i.e., patterns of links, (see different methods for the community detection in binary graphs in \cite{danon2006,newman2003}), but also according to the weights of the links, which are not necessarily associated with the topology.   Recently different methods were adjusted to detecting communities in the weighted networks, such as weighted maximum-likelihood-method (wMLM) \cite{mitrovic2008} and spectral analysis of the adjacency or other weighted matrices related to it \cite{donetti2004,mitrovic2008b}. Here we apply the spectral methods to determine communities in the  weighted user subgraph.  

{\it Spectral methods to find communities} in modular networks is based on the properties of the eigenvalues and eigenvectors of the adjacency matrix or other matrix related to it.  We use weighted normalized Laplacian (diffusion operator) \cite{mitrovic2008b,samukhin2007}  
\begin{equation}
L_{ij}^U=\delta_{ij}-\frac{C_{ij}^B}{\sqrt{l_{i}l_{j}}}  \ '
\label{eq-Lap}
\end{equation}
where $C_{ij}^B$ are weighted links determined by the number of common posts and comments for the user pair $\{ij\}$, and 
$l_{i}$ is the strength of the user node $i$, standardly defined as the  sum of weights $\ell _i \equiv \sum _jC_{ij}^B$ of all links at that node. Note that with the projection of our bipartite graph onto monopartite (user partition) the directions of the links are lost, thus we consider the links of the projected graph as a symmetrical.

\begin{figure}
\centering
\begin{tabular}{cc}
\resizebox{18pc}{!}{\includegraphics{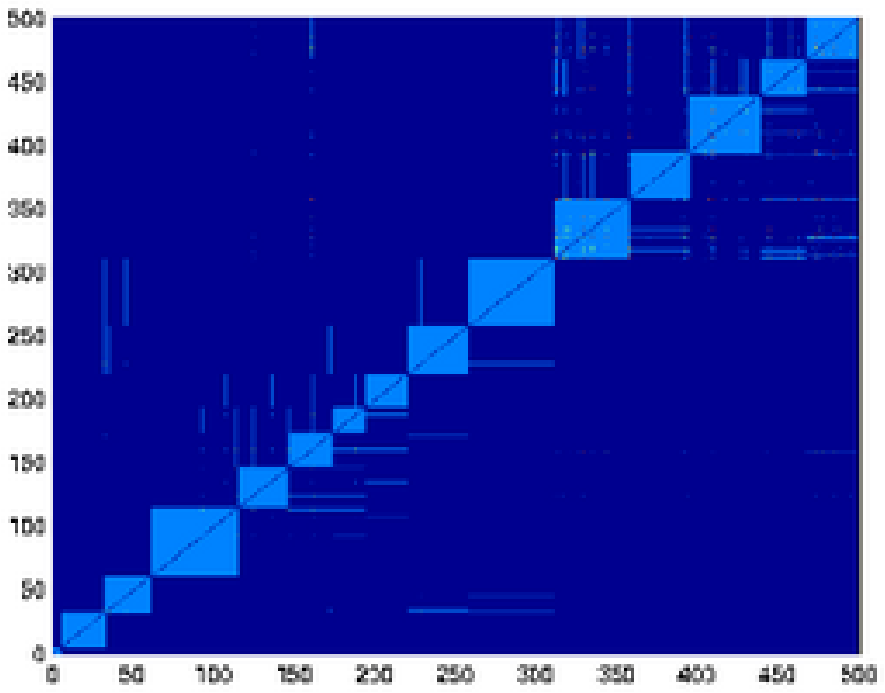}}\\
\resizebox{20pc}{!}{\includegraphics{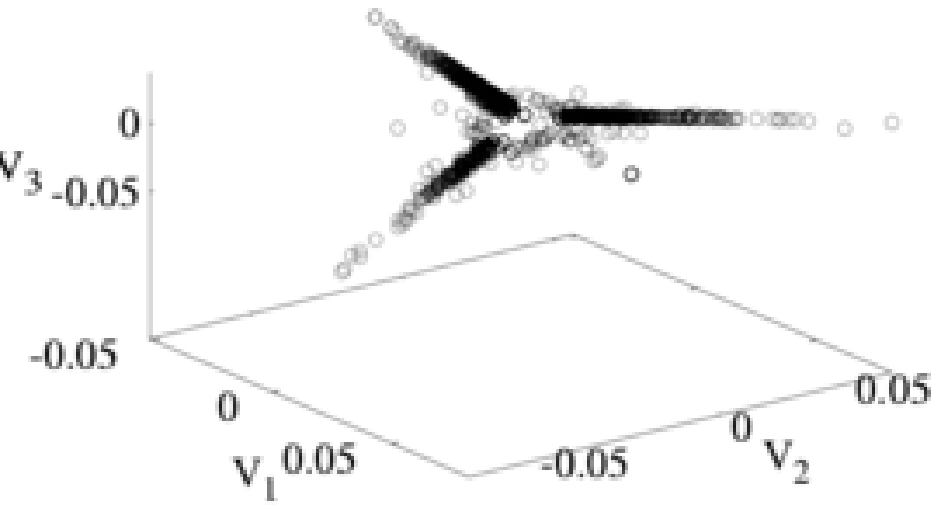}}
\end{tabular}
\caption{BBC Blogs: Part of the weighted matrix of commons $C^B_{ij}$ (top)  and the scatter plot of three eigenvectors of the Laplacian (\ref{eq-Lap}) of small  nonzero eigenvalues, exhibiting three user communities (bottom).}
\label{fig-bbcscatter-plot}
\end{figure}

As discussed in detail in Ref.\ \cite{mitrovic2008b}, the normalized Laplacian in the form (\ref{eq-Lap}) has the spectrum limited in the range $[0,2]$ and the orthogonal set of eigenvectors. The essence of the community detection  by spectral methods consists in the following observations:
\begin{itemize}
\item Lowest non-zero eigenvalues appear to be separated from the rest of the spectrum; Their number coincides with the number of distinguishable subgraphs;
\item Eigenvectors of these lowest non-zero eigenvalues  have non-zero positive/negative components localized on the subgraphs;
\item Scatter-plot in the space of these eigenvectors exhibits branched structure; Points in each branch carry the index of a non-zero component, which uniquely indicates a node in the network; Thus, well separated branches of the scatter-plot identify different subgraphs on the network. 
\end{itemize}
We expect to find a community structure, for instance in the case of BBC Blogs due to the predetermined category of Blogs. However, it is not entirely clear how this structure of Blogs affects user behavior. In Fig.\ \ref{fig-bbcscatter-plot}(top) we show a part of the weighted matrix $C_{ij}^B$ obtained from the BBC Blogs.  Small and larger blocks along the diagonal of the matrix indicate linking between subgroups of posts, which is achieved via  users of these posts. We analysed the Laplacian matrix for the respective projection on the user partition of the BBC Blogs, as described above, Eq. (\ref{eq-Lap}).  The resulting scatter-plot is shown in Fig.\ \ref{fig-bbcscatter-plot}(bottom). In this section of the eigenvector space, the scatter-plot exhibits  three well separated groups of users, represented by large branches, and the central ring. Matching the identity of users at tips of these three branches, we then identify the lists of posts  which thy commented. Indeed, these three  users groups appear to be related to three different subjects of posts (Sports, Business and economy, and Technology Blogs). Forth group is rather small and it is also related to  Sports posts. Note that here we considered only users related to posts with the number of comments between 50 and 100, which gives $N_{npB}= 149$  posts and $N_U=4957$ users. The other group of very popular posts on BBC Blogs will be discussed later.

In the case of B92 Blogs no predefined category of Blogs exists, and  a community structure may appear in a self-organized manner.  We first consider a group of posts  with less than $100$ comments (corresponding to out-degree $q^B<100$, roughly the banding point in Fig.\ \ref{fig-degree}, bottom). The selected set consists of $N_{npP}=3318$ posts. These posts together with corresponding comments constitute subgraph of size $N_{npB}=137941$, commented by $N_{npU}=3367$ users. Bipartite network constructed from non-popular group is then projected in the way described in the previous section and the monopartite weighted network of  $N_{npU}$ users is determined and its spectra analysed.

\begin{figure}
\centering
\begin{tabular}{cc}
\resizebox{18pc}{!}{\includegraphics{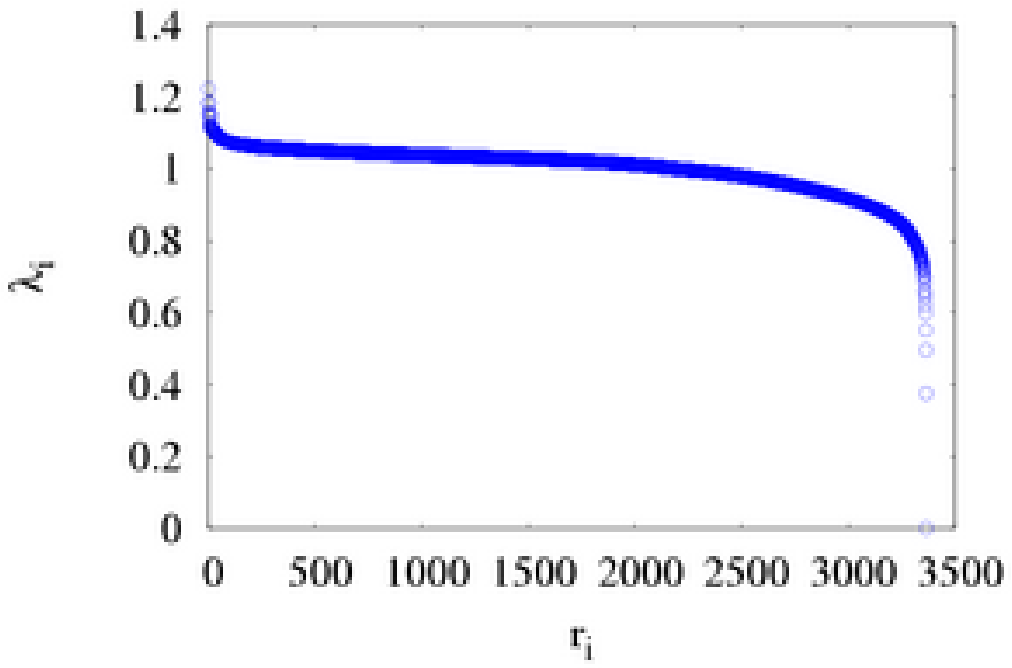}}\\
\resizebox{20pc}{!}{\includegraphics{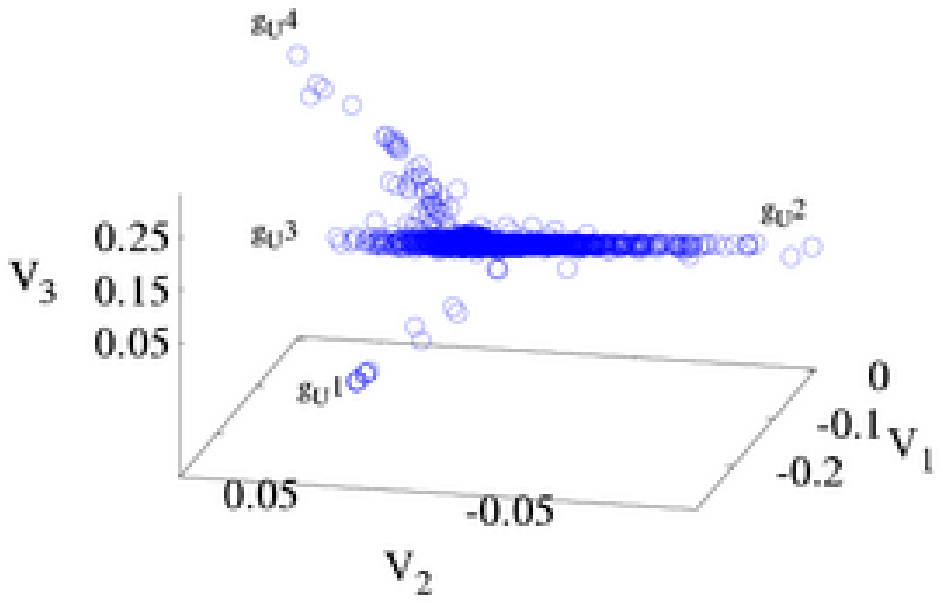}}
\end{tabular}
\caption{Eigenvalues in ranking order (top) and the eigenvectors of three lowest non-zero eigenvalues (bottom) of the weighted Laplacian matrix of the network of users on B92 Blogs. Four groups of users are marked.
}
\label{fig-B92scatter_plot}
\end{figure}

Ranking of the eigenvalues of the respective Laplacian matrix and the scatter-plot in the space of three representative eigenvectors is shown in Fig.\ \ref{fig-B92scatter_plot}. Four groups (user communities) are clearly differentiated in the spectrum, and are marked as $ g_U1$, $g_U2$, $g_U3$, $g_U4$. In the following we analyse the structure of Blogs to which these four user groups are related.

\begin{figure}
\centering
\begin{tabular}{cc}
\resizebox{16.2pc}{!}{\includegraphics{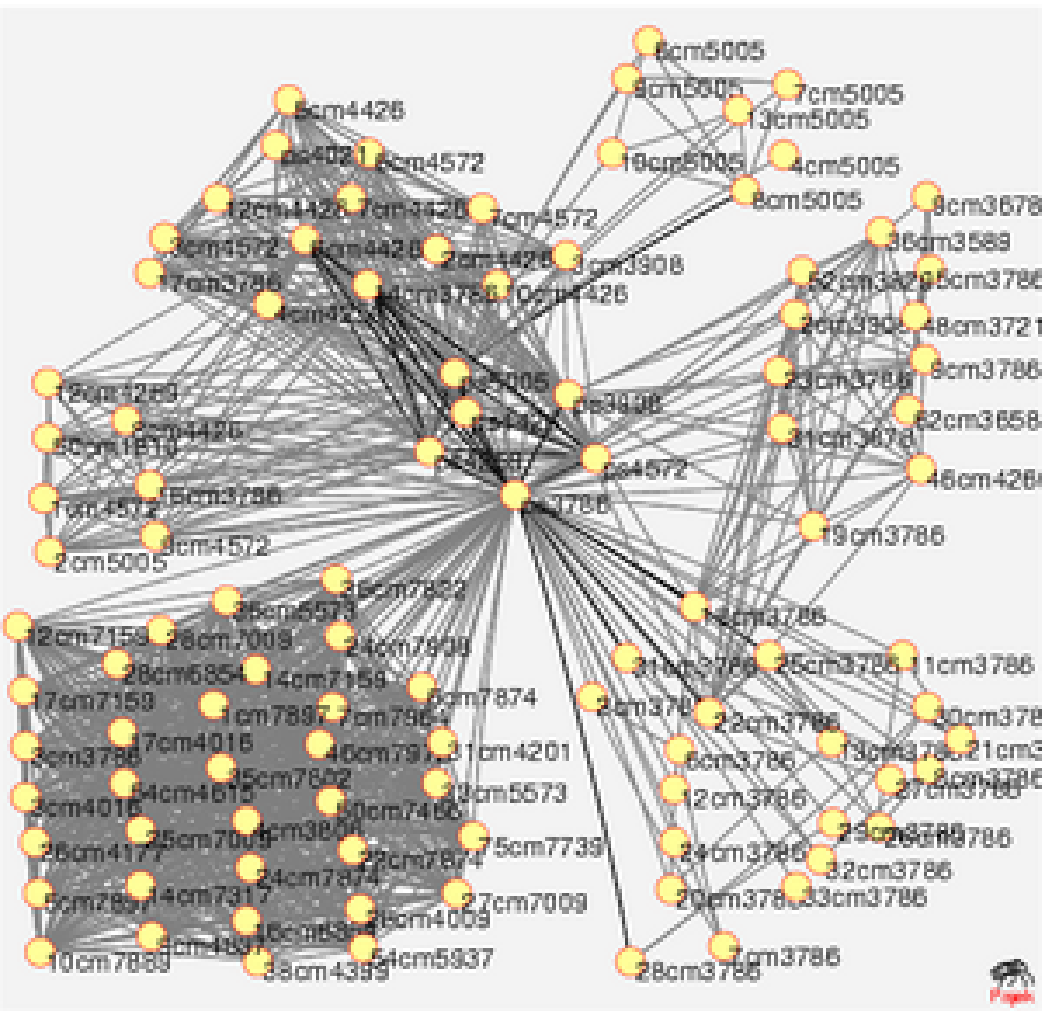}}\\
\resizebox{16pc}{!}{\includegraphics{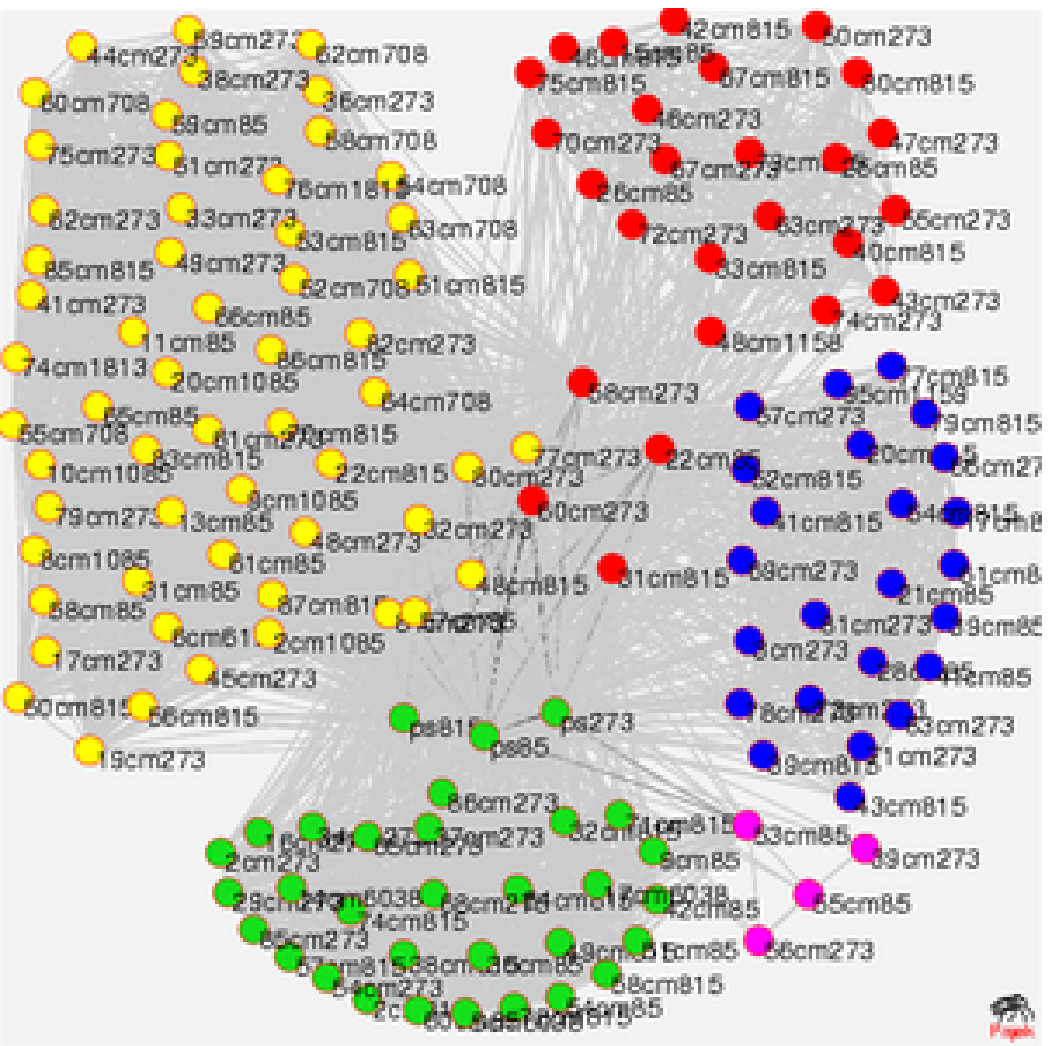}}\\
\end{tabular}
\caption{The network of posts and comments linked to groups of users $g_U1$ (top ) $g_U4$ (bottom) can be further split into communities, which are related either to the subject or to the author of the post.}
\label{fig-postsug1ug4}
\end{figure}

The   users in groups $ g_U1$ and $g_U4$ (lower and upper branch in Fig.\ 
\ref{fig-B92scatter_plot}), are linked to the Posts which we show as the network Fig.\ \ref{fig-postsug1ug4}. By inspection of the text in these posts and related comments, we find that all posts in the user group $ g_U1$  are related to sports (football), and posts in group $g_U4$ are about the urban architecture and related urban life problems [Tables 1,2 of these posts are available on Supportive material, in Serbian]. In Fig.\ \ref{fig-postsug1ug4} these posts are further differentiated according to the authors of the posts or comments.  Although these two groups are rather small, one can see that the number of posts and comments by different users varies, in accordance with the broad distribution of the user degree on the bipartite network, mentioned in sec.\ \ref{sec-binet}.

\begin{figure}
\centering
\begin{tabular}{cc}
\resizebox{20pc}{!}{\includegraphics{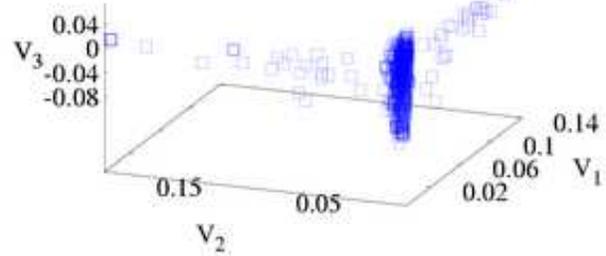}}\\
\resizebox{18pc}{!}{\includegraphics{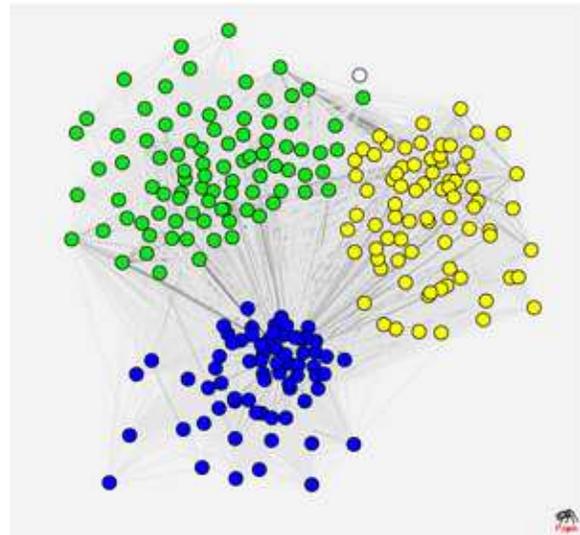}}
\end{tabular}
\caption{(top) Scatter plot of three eigenvectors of the network of posts linked to the user group $g_U2$, recent posts. In the wings are the  posts written by two different authors. (bottom) Network of early blogs related to user group $g_U3$. Three groups are distinguished by using the weighted wMLM. }
\label{fig-gu2blogs-scatter}
\end{figure}

In contrast to the above thematic posts, the posts and comments related to the other two user groups $g_U2$ and  $g_U3$, are on mixed subjects and involve a larger community of users, however, they share more similarities apart from excluding the sports and the architecture subjects. The difference in posts related to both ends of these two branches is the time when they appeared: that posts related to the users on the left end of the scatter plot ($g_U3$) are posted at the beginning of the Blog site, as opposed to the right branch ( $g_U2$), which are related to the recent posts. As above, we can further analyse the structure of subnetwork of posts related to the user group  $g_U2$ and $g_U3$, see Fig.\ \ref{fig-gu2blogs-scatter} and the caption to the Figure for details.

\subsection{Structure of Communities in Popular Blogs}
\begin{figure}
\centering
\begin{tabular}{cc}
\resizebox{24pc}{!}{\includegraphics{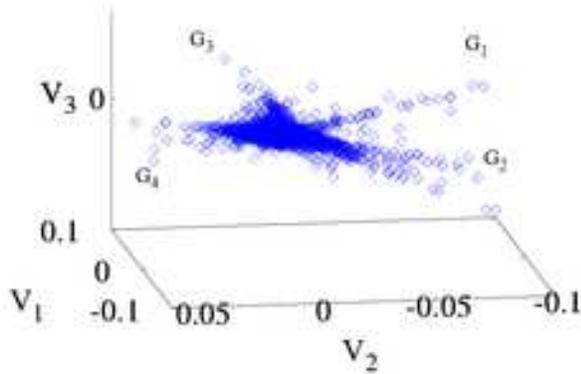}}\\
\end{tabular}
\caption{Community structure in the bipartite network of users + popular posts from the data of B92 Blogs.}
\label{fig-popularB92scatter}
\end{figure}

\begin{figure}[h]
\centering
\begin{tabular}{cc}
\resizebox{22pc}{!}{\includegraphics{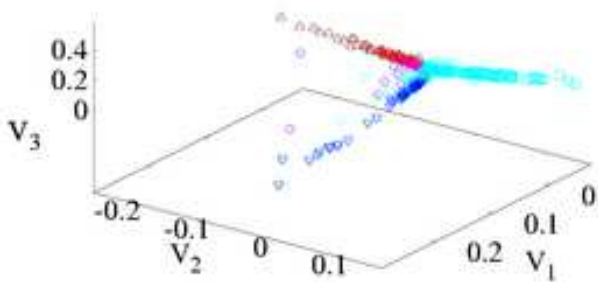}}\\
\end{tabular}
\caption{Further splitting in the structure of groups $G_1+G_2$ from Fig.\ \ref{fig-popularB92scatter}.}
\label{fig-popularB92-G1G2}
\end{figure}

In the case of popular Blogs, we first filter the data according to the number of comments that exceeds 100 on a post and users who wrote these posts and the related comments. We then construct a bipartite network consisting of Users + Posts, while the comments are used to define the weights of links between Users and Posts. Namely, the weight $W_{ij}$ of the link between the user $i$ and post $j$ is defined by the number of comments that user $i$ left on the post $j$.  The network of Users + Posts consists of $N= 5079$ nodes (1466 posts and 3613 users). The weighted Laplacian matrix of the whole network is constructed as in Eq.\ (\ref{eq-Lap}) but with the $W_{ij}$ weights and its spectrum computed. The structure of communities in this network is shown in the 3-dimensional scatter-plot of the eigenvectors in Fig.\ \ref{fig-popularB92scatter}. In this projection four branches of nodes can be differentiated, denoted as $G_i$, $=1,2,3,4$. Note that each point in this scatter plot is either a User (known by its IDu) or a Post (known as its IDb). In order to separate the users from posts (two partitions in our bipartite network) one can, for instance, apply the wMLM metod to each of the observed groups $G_i$.
 Here we are interested in the contents of the popular Blogs. For this purpose we further break  the observed groups according to the spectrally detectable communities. We demonstrate  it on the example of the network made of the groups $G_1+G_2$ from the above Fig.\ \ref{fig-popularB92scatter}.  The corresponding structure of the communities is shown in Fig.\ \ref{fig-popularB92-G1G2}. 

By inspection of IDb and the text on these posts, we find that the group $G_1$ of the previous plot remains the same (it is related to the Montenegrian political issues), while the group $G_2$ splits into three groups. A small group appears in the vertical plane contains posts related to internal politics. In the other two branches are the posts related to the rights of pregnant women in hospitals in Serbia (left branch) and other still mixed issues and many users (right branch). For fine structure of posts in these branch one needs to include additional criteria, e.g., related to the contents of the text in these posts.
We performed such analysis on popular BBC Blogs, the results are left out of this paper.

\section{Conclusions\label{sec-Summary}}
We have presented analysis of large  Blog datasets, seen from the point of view of complex dynamical systems and represented by bipartite graphs of Users and their Posts and Comments. We studied in parallel data sets from two Blog sites with entirely different history, cultural, and organizational profiles. Nevertheless, the statistical features appear to be surprisingly similar and robust.

In this paper we have focused on  two aspects:
\begin{itemize}
\item Evolution of the activity on Blogs leading to the robust temporal patterns. The similarity with some other social media \cite{crane2009} are apparent and may be eventually attributed to human nature and preferences;
\item Emergence of user communities in the Blogs-mediated interactions. On Blogs the users appear to be normally clustered around preferred subjects. However, mixing between subjects increases when a post becomes popular, suggesting another active principle, which can be found by a direct inspection of the posted material;  
\end{itemize}
Although the underlying mechanisms that drive the user activity on Blogs are not exactly known, our results suggest that the data alone, traced over time and studied by the appropriate network methods, reveal a lot of structure in the Blog space.  The community  analysis as we presented it here, can be used as a staring point to further effective study of the contents on Blogs, which need to be related with clusters of user rather than individual users. Furthermore, by unraveling the  user profiles (i.e., from the history of their actions stored under their registered IDs), the results of our study may be useful  for developing appropriate theoretical models, which are necessary to eventually understand the key mechanisms in the Blog dynamics. These two aspects remain for the future study.

{\bf Acknowledgments} Research supported  by the program  P1-0044
(Slovenia) and by EU FP7 project CYBEREMOTIONS.
The numerical results were obtained on the computer grid of the Department of theoretical physics at Jo\v zef Stefan Institute, Ljubljana, and  on the 
AEGIS e-Infrastructure at the Institute of Physics, Belgrade, supported in part by EU FP6 and FP7 projects EGEE-III, SEE-GRID-SCI and CX-CMCS.
\bibliographystyle{unsrt}
\bibliography{blog.bib}

\begin{thebibliography}{10}

\bibitem{nature09}
G.~{Brumfiel}.
\newblock {Breaking the convention?}
\newblock {\em Nature}, 459:1050--1051, 2009.

\bibitem{grujic2009}
J.~{Gruji\'c}, M.~{Mitrovi\'c}, and B.~{Tadi\'c}.
\newblock Mixing patterns and and communities on bipartite graphs on web-based
  social interactions.
\newblock {\em IEEEXplore}, 2009.

\bibitem{lorenz2008}
J.~Lorenz.
\newblock Universality of movie rating distributions.
\newblock {\em arXiv:0806.2305}, 2008.

\bibitem{lambiotte2005}
R.~Lambiotte and M.~Ausloos.
\newblock Uncovering collective listening habits and music genres in bipartite
  networks.
\newblock {\em Physical Review E}, 72:066107, 2005.

\bibitem{lambiotte2006}
R.~Lambiotte and M.~Ausloos.
\newblock On the genre-fication of music: a percolation approach (long
  version).
\newblock {\em The European Physical Journal B}, 50:183, 2006.

\bibitem{kujawski2007}
B.~{Kujawski}, J.~{Holyst}, and G.~J. {Rodgers}.
\newblock {Growing trees in internet news groups and forums}.
\newblock {\em Physicsal Review E}, 76:036103, 2007.

\bibitem{bachnik2005}
W.~{Bachnik}, S.~{Szymczyk}, S.~{Leszczynski}, R.~{Podsiadlo}, E.~{Rymszewicz},
  L.~{Kurylo}, D.~{Makowiec}, and B.~{Bykowska}.
\newblock {Quantitive and Sociological Analysis of Blog Networks}.
\newblock {\em Acta Physica Polonica B}, 36:3179--+, 2005.

\bibitem{fu2006}
F.~{Fu}, L.~{Liu}, K.~{Yang}, and L.~{Wang}.
\newblock {The structure of self-organized blogosphere}.
\newblock {\em arXiv:0607361}, 2006.

\bibitem{leskovec2007}
J.~{Leskovec}, M.~{McGlohon}, C.~{Faloutsos}, N.~{Glance}, and M.~{Hurst}.
\newblock {Cascading Behavior in Large Blog Graphs}.
\newblock {\em arXiv:0704.2803}, 2007.

\bibitem{liu2007}
L.~{Liu}, F.~{Fu}, and L.~{Wang}.
\newblock {Information propagation and collective consensus in blogosphere: a
  game-theoretical approach}.
\newblock {\em arXiv:0701316}, 2007.

\bibitem{sano2009}
Y.~Sano and M.~Takayasu.
\newblock Macroscopic and microscopic statistical properties observed in blog
  entries.
\newblock {\em arXiv:0906.1744}, 2009.

\bibitem{mike2007}
M.~{Thelwall}, A.~{Byrne}, and M.~{Goody}.
\newblock {Which types of news story attract bloggers?}
\newblock {\em Informationresearch}, 12:327, 2007.

\bibitem{thelwall2009b}
M.~{Thelwall}, D.~{Wilkinson}, and S.~{Uppal}.
\newblock {Data Mining Emotion in Social Network Communication: Gender
  differences in MySpace}.
\newblock {\em Journal of the American Society for Information Science and
  Technology}, 2009.

\bibitem{fortuna2007}
B.~{Fortuna}, M.~{Grobelnik}, and D.~{Mladenic}.
\newblock Ontogen.
\newblock http://ontogen.ijs.si/, 2008.

\bibitem{science2009}
A.~{Cho}.
\newblock {Ourselves and Our Interactions: The Ultimate Physics Problem?}
\newblock {\em Science}, 325, 2009.

\bibitem{eisler2008}
Z.~{Eisler} and J.~{Kert\'esz}.
\newblock Random walks on complex networks with inhomogeneous impact.
\newblock {\em Physical Review E}, 71(5):057104, 2005.

\bibitem{zivkovic2006}
J.~{{\v Z}ivkovi{\'c}}, B.~{Tadi{\'c}}, N.~{Wick}, and S.~{Thurner}.
\newblock {Statistical indicators of collective behavior and functional
  clusters in gene networks of yeast}.
\newblock {\em European Physical Journal B}, 50:255--258, 2006.

\bibitem{crane2009}
R.~{Crane}, F.~{Schweitzer}, and D.~{Sornette}.
\newblock {New Power Law Signature of Media Exposure in Human Response Waiting
  Time Distributions}.
\newblock {\em arXiv:0903.1406}, 2009.

\bibitem{grinstein2008}
G.~{Grinstein} and R.~{Linsker}.
\newblock {Power-law and exponential tails in a stochastic priority-based model
  queue}.
\newblock {\em Physical Review E}, 77(1):012101--+, 2008.

\bibitem{tadic2004}
Bosiljka Tadi\ifmmode~\acute{c}\else \'{c}\fi{}, Stefan Thurner, and G.~J.
  Rodgers.
\newblock Traffic on complex networks: Towards understanding global statistical
  properties from microscopic density fluctuations.
\newblock {\em Phys. Rev. E}, 69(3):036102, Mar 2004.

\bibitem{grujic2008}
J.~{Gruji\'c}.
\newblock Movies recommendation networks as bipartite graphs.
\newblock {\em Lecture Notes in Computer Science}, 5102, 2008.

\bibitem{danon2006}
L.~{Danon}, A.~{D\'{i}az-Guilera}, and A.~{Arenas}.
\newblock The effect of size heterogeneity on community identification in
  complex networks.
\newblock {\em Journal of Statistical Mechanics: Theory and Experiment},
  11:P11010, 2006.

\bibitem{newman2003}
M.~E.~J. Newman.
\newblock Mixing patterns in networks.
\newblock {\em Phys. Rev. E}, 67:026126, 2003.

\bibitem{mitrovic2008}
M.~{Mitrovi\'c} and B.~{Tadi\'c}.
\newblock {Search of Weighted Subgraphs on Complex Networks with Maximum
  Likelihood Methods}.
\newblock {\em Lecture Notes in Computer Science}, 5102:551--558, 2008.

\bibitem{donetti2004}
L.~{Donetti} and M.~A. {Mu{\~n}oz}.
\newblock {Detecting network communities: a new systematic and efficient
  algorithm}.
\newblock {\em Journal of Statistical Mechanics: Theory and Experiment}, 10,
  2004.

\bibitem{mitrovic2008b}
M.~{Mitrovi\'c} and B.~{Tadi\'c}.
\newblock Spectral and dynamical properties in classes of sparse networks with
  mesoscopic inhomogeneities.
\newblock {\em Physical Review E}, 80, 2009.

\bibitem{samukhin2007}
A.~N. {Samukhin}, S.~N. {Dorogovtsev}, and J.~F.~F. {Mendes}.
\newblock {Laplacian spectra of, and random walks on, complex networks: Are
  scale-free architectures really important?}
\newblock {\em Physical Review E}, 77(3):036115--+, 2008.

\end{thebibliography}
\end{document}